\documentclass[conference,final,10pt]{IEEEtran}
\usepackage[left=0.625in, right=0.625in, bottom=1.1in, top=0.71in]{geometry}
\usepackage[utf8]{inputenc}
\usepackage{graphicx}
\usepackage{amsmath,amssymb,amsthm,mathtools}
\usepackage{color}
\usepackage{upgreek}
\usepackage{ragged2e}
\usepackage{subfigure}
\usepackage{stfloats}
\usepackage{float}
\usepackage{bm}
\usepackage{cite}
\usepackage{cases}
\usepackage{subfigure}
\usepackage{hyperref}
\usepackage{mathrsfs}
\usepackage{multicol}
\usepackage{algorithm, algpseudocode}

\usepackage[acronym,shortcuts]{glossaries}

\algrenewcommand\algorithmicindent{1em}
\algnewcommand\algorithmicforeach{\textbf{for each}}
\algdef{S}[FOR]{ForEach}[1]{\algorithmicforeach\ #1\ \algorithmicdo}

     \makeatletter
    \def\footnoterule{\kern-2\p@
      \noindent\hrulefill \kern 2.6\p@ \vspace{0.5ex}} % the \hrule is .4pt high
    \makeatother

\newacronym{5G}{5G}{fifth generation}
\newacronym{eMBB}{eMBB}{enhanced mobile broadband}
\newacronym{mMTC}{mMTC}{massive machine type communication}
\newacronym{URLLC}{URLLC}{ultra-reliable low latency communication}
\newacronym{IoT}{IoT}{internet of things}
\newacronym{MIMO}{MIMO}{multiple-input multiple-output}
\newacronym{DoF}{DoF}{degree of freedom}
\newacronym{LRMC}{LRMC}{low-rank matrix completion}
\newacronym{MSE}{MSE}{mean square error}
\newacronym{MMSE}{MMSE}{minimum mean square error}
\newacronym{NMSE}{NMSE}{normalized mean square error}
\newacronym{MC}{MC}{matrix completion}
\newacronym{NP}{NP}{non-deterministic polynomial-time}
\newacronym{CS}{CS}{compressed sensing}
\newacronym{SDP}{SDP}{semidefinite program}
\newacronym{RAM}{RAM}{random access memory}
\newacronym{TNN}{TNN}{truncated nuclear norm}
\newacronym{NN}{NN}{nuclear norm}
\newacronym{nmAPG}{nmAPG}{nonmonotone accelerated proximal gradient}
\newacronym{niAPG}{niAPG}{nonconvex inexact accelerated proximal gradient}
\newacronym{PG}{PG}{proximal gradient}
\newacronym{SVT}{SVT}{singular value thresholding}
\newacronym{LSP}{LSP}{log-sum-penalty}
\newacronym{FD}{FD}{full-duplex}
\newacronym{HD}{HD}{half-duplex}
\newacronym{RF}{RF}{radio-frequency}
\newacronym{SI}{SI}{self-interference}
\newacronym{PA}{PA}{power amplifier}
\newacronym{DAC}{DAC}{digital-to-analog converter}
\newacronym{ADC}{ADC}{analog-to-digital converter}
\newacronym{AWGN}{AWGN}{additive white Gaussian noise}
\newacronym{CSI}{CSI}{channel state information}
\newacronym{LoS}{LoS}{line-of-sight}
\newacronym{NLoS}{NLoS}{non line-of-sight}
\newacronym{i.i.d.}{i.i.d.}{independent and identically distributed}
\newacronym{AoD}{AoD}{angle of departure}
\newacronym{AoA}{AoA}{angle of arrival}
\newacronym{ULA}{ULA}{uniform linear array}
\newacronym{SINR}{SINR}{signal-to-interference-plus-noise ratio}
\newacronym{LNA}{LNA}{low noise amplifier}
\newacronym{QCQP}{QCQP}{quadratic constrained quadratic program}

\renewcommand{\smallskip}{\vspace{0.25cm}}
\newcommand{\Exp}[1]{\mathbb{E}\left[{#1}\right]}
\newcommand{\norm}[1]{\left\lVert#1\right\rVert}

\newcommand{\tr}[1]{{\rm Tr}\left({#1}\right)}
\newcommand{\diag}[1]{{\rm diag}\!\left({#1}\right)\!}

\newcommand{\argmin}{\mathop{\rm argmin}\limits}
\newcommand{\argmax}{\mathop{\rm argmax}\limits}
\renewcommand\footnoterule{{\hrule height 0pt}}

\begin{document}

\title{Full-Duplex MIMO Systems with Hardware Limitations and Imperfect Channel Estimation}

\author{
\IEEEauthorblockN{Hiroki Iimori$^\dagger$, Giuseppe Thadeu Freitas de Abreu$^{\dagger}$, and Koji Ishibashi$^{*}$\\}\vspace{1ex}
\IEEEauthorblockA{
$^\dagger$ Department of Comp. Sci. and Elec. Eng., Jacobs University Bremen, Campus Ring 1, 28759, Bremen, Germany \\
$^{*}$ AWCC, The University of Electro-Communications,
1-5-1 Chofugaoka, Chofu-shi, Tokyo 182-8585, Japan}
}

\maketitle

\begin{abstract}
We consider a bidirectional in-band \ac{FD} \ac{MIMO} system subject to imperfect \ac{CSI}, hardware distortion, and limited analog cancellation capability as well as the \ac{SI} power requirement at the receiver analog domain so as to avoid the saturation of \ac{LNA}.
A novel \ac{MMSE}-based joint design of digital precoder and combiner for \ac{SI} cancellation is offered, which combines the well-known gradient projection method and non-monotonicity considered in recent machine-learning literature in order to tackle the non-convexity of the optimization problem formulated in this article.
Simulation results illustrate the effectiveness of the proposed \ac{SI} cancellation algorithm.
\end{abstract}
\glsresetall

% Introduction ------------------------------------
\vspace{-1ex}
\section{Introduction}
\label{sect:intro}
\vspace{-0.5ex}

With the beginning of the \ac{5G} era architected to support  individual service categories, in particular \ac{eMBB}, \ac{mMTC}, and \ac{URLLC}), in-band \ac{FD} technology, which enables simultaneous transmission and reception on the same time-frequency resource block, has been considered as a promising alternative to its \ac{HD} counterpart, as it can be leveraged to jointly tackle different system requirements such as overhead reduction, resource scarcity problem, and demands for higher data rates.

Despite the fact that the concept of \ac{FD} communications was developed decades ago, wireless \ac{FD} operation has -- due to the overwhelming \ac{SI} caused by leakage of its own transmitted signals which results from the close proximity between transmit and receive antennas installed on the \ac{FD} radio -- long been considered infeasible in practice until experimental and theoretical research work demonstrating otherwise emerged in the beginning of the 2010s \cite{bharadia2013fullduplex, Everett2011Asilomar, Choi2010MobiCom, duarte2010fullduplex, Jain2011MobiCom, Bliss2007WSSP, RadunovicWIMESH2010}. 
Motivated by the above, a substantial amount of research contributions to \ac{SI} cancellation technology in conjunction with \ac{MIMO} for higher spatial \acp{DoF}  has been amassed \cite{vehkapera2013asymptotic,Iimori2018Globalsip,DaySP2012,Riihonen2013CISS,GowdaTWC2018,MyListOfPapers:KorpiGlobecom2014,IimoriSPAWC2018}, demonstrating theoretical feasibility of the in-band \ac{FD} operation under the assumption that ideal \ac{CSI} knowledge and/or \ac{RF} hardware architectures are available.
To mention a few examples, the authors in \cite{MyListOfPapers:KorpiGlobecom2014} have studied a hybrid analog-digital \ac{SI} cancellation architecture for \ac{FD} \ac{MIMO} systems with fully-connected analog cancellation taps, whereas \cite{OmidWSA15} investigated an interference mitigation scheme aiming at not only the \ac{SI} but also inter-user interference in a multi-cell multi-user scenario.

However, the performance of \ac{SI} cancellation mechanisms for \ac{FD} is bounded not only by channel estimation inaccuracy but also by non-ideal hardware distortions including nonlinearity of \acp{PA}, \acp{DAC}, and I/Q mixers, leading to the necessity of incorporating such imperfections into the design of \ac{SI} cancellation \cite{OmidTVT18}.
To make matters worse, it has been argued recently \cite{GowdaTWC2018, MyListOfPapers:KorpiGlobecom2014, Sim2017CM, Cirik2013Asilomar, Kolodziej2016TWC, Alexandropoulos2017, Liu2017,AnhTWC20} that the architectural and computational complexity as well as the associated energy consumption in order to perform these hybrid digital-analog \ac{SI} cancellation will be prohibitive as the number of antennas increases, imposing a new challenge on \ac{SI} cancellation under limited analog cancellation capability.

In order to tackle this difficulty, a low-complexity \ac{SI} cancellation method subject to limited analog cancellation capability for large-scale \ac{FD} \ac{MIMO} systems was proposed in \cite{GowdaTWC2018}, and a new analog cancellation architecture based on tap delay line processing such that the number of analog cancellation taps can be reduced  while maintaining the spatial \ac{DoF} for the desired system performance was introduced in \cite{Kolodziej2016TWC}.
Leveraging the latter, \cite{Alexandropoulos2017} studied a \ac{FD} \ac{MIMO} system equipped with the low-complexity multi-tap analog canceller proposed in \cite{Kolodziej2016TWC} under the assumption of perfect \ac{CSI} and ideal hardware components, which further extended in \cite{Liu2017} to an imperfect \ac{CSI} scenario without considering hardware distortion.
Aiming to simultaneously take into account hardware impairments, imperfect \ac{CSI} and limited hardware complexity for analog \ac{SI} cancellation, the authors in \cite{IimoriTWC19} proposed a low-complexity spatial-temporal \ac{SI} cancellation design for bidirectional \ac{FD} \ac{MIMO} systems.

One of bottlenecks of contributions such as the ones mentioned above is, however, that the \ac{SI} power level at the receiver analog domain is not properly tuned so as to avoid the saturation of the \ac{LNA}, which is still a major challenge to be conquered.
In this paper, we therefore propose an algorithmic solution to the latter problem for bidirectional \ac{FD} \ac{MIMO} communications, while taking all the aforementioned issues ($i.e.,$ imperfect \ac{CSI}, hardware distortion, and limited analog cancellation capability) into consideration.

The remainder of the article is as follows.
In Section \ref{sect:system_model}, the system model including imperfect \ac{CSI} and hardware distortion is given, where \ac{SINR} expressions and the \ac{SI} power at receiver analog domain are also mathematically described.
The problem formulation for the desired \ac{SI} cancellation will be discussed in Section \ref{sec:proposed}, in which the proposed gradient projection based \ac{SI} cancellation design is also offered.
In Section \ref{sec:results}, as an illustration, simulation results are given in order to demonstrate the effectiveness of the proposed method.
Finally, conclusions and discussions on possible future works are given in Section \ref{sec:conclusion}.

\emph{Notation:}
Throughout the article, matrices and vectors will be expressed respectively by bold capital and small letters, namely, $\bm{X}$ and $\bm{x}$.
The transpose, conjugate, Hermitian and inverse operators will be respectively denoted by $\left(\cdot\right)^\mathrm{T}$, $\left(\cdot\right)^{*}$, $\left(\cdot\right)^\mathrm{H}$ and $\left(\cdot\right)^{\footnotesize-1}$, while the expectation, the covariance and the Frobenius norm operators will be respectively denoted by $\Exp{\cdot}$, $\mathbb{V}\left[\cdot\right]$ and $\norm{\cdot}$.
A complex matrix with $a$ columns and $b$ rows is denoted by $\bm{X}\in\mathbb{C}^{a\times b}$, and a complex random scalar variable following the complex Gaussian distribution with mean $\mu$ and variance $\sigma^2$ is expressed as $x \sim \mathcal{CN}\left(\mu,\sigma^2\right)$.
Finally, the matrix containing only the diagonal of $\bm{X}$ will be denoted by $\diag{\!\bm{X}\!}$\,.

% Model ------------------------------------
\section{System Model}
\label{sect:system_model}

\begin{figure}[H]
\center
\includegraphics[width=\columnwidth]{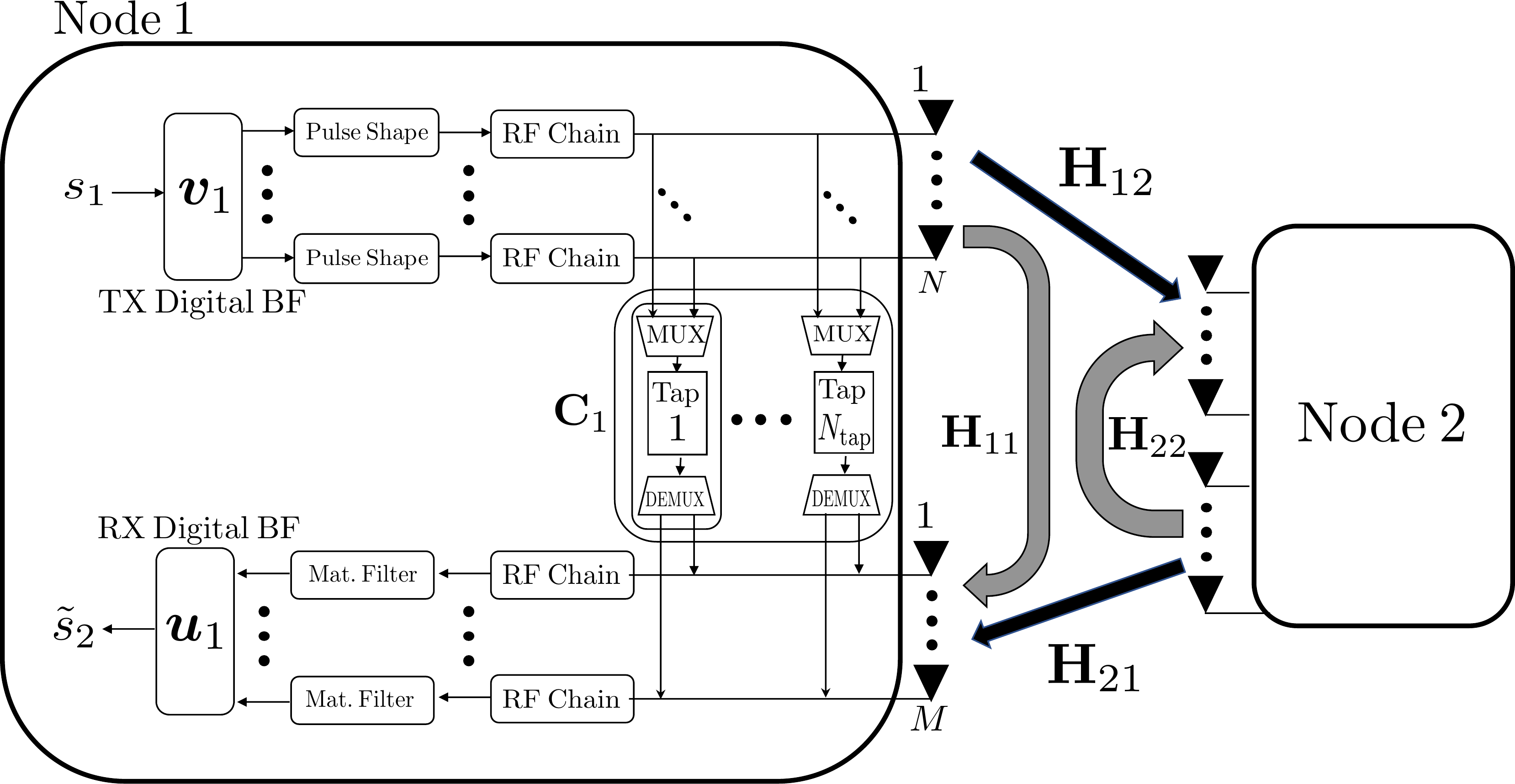}\vspace{-1ex}
\caption[]{Two-way in-band \ac{FD} \ac{MIMO} system model equipped with digital precoder/combiner and multi-tap analog canceller.}
\label{fig:System_model}
\end{figure}

Consider a bidirectional two-way in-band \ac{FD} \ac{MIMO} system shown in Figure \ref{fig:System_model}, where two nodes operating in \ac{FD} mode exchange information with the support of a digital precoding vector $\bm{v}_k\in\mathbb{C}^{N\times 1}$ with $k\in\{1,2\}$, a digital combining vector $\bm{u}_{k}\in\mathbb{C}^{1\times M}$, and a low-complexity multi-tap analog \ac{SI} cancellation architecture \cite{Kolodziej2016TWC}, such that each node is capable of suppressing the \ac{SI} while increasing the intended signal power at the destination node.
For the sake of simplicity but loss of generality, each node is assumed to be equipped with $N$ transmit and $M$ receive antennas, respectively.
Due to the limited dynamic range of \ac{RF} components at the nodes, it is assumed that each node suffer from not only inevitable \ac{SI} caused by own transmitted signals but also nonlinear hardware impairments from non-ideal \acp{PA}, \acp{DAC} and I/Q mixer.

It is further assumed that the transmit power at the $k$-th node is limited \big($i.e.,$ $\Exp{\norm{\bm{v}_{k}s_{k}}^2}\leq P_{k}$\big) with $s_{k}$ denoting a unit-power symbol transmitted by the $k$-th node, whereas the combining vector $\bm{u}_{k}$ is normalized ($i.e.,$ $\norm{\bm{u}_{k}}^2=1$).
Following \cite{Kolodziej2016TWC,IimoriTWC19,Iimori2018Globalsip}, the low-complexity multi-tap analog \ac{SI} cancellation at the $k$-th node can be expressed as $\bm{C}_{k}\in\mathbb{C}^{M\times N}$ composed of $N_{\rm tap}$ non-zero components and $MN - N_{\rm tap}$ zeros. 

Referring to Figure \ref{fig:System_model}, the communication channel from the $\ell$-th node, with $\ell\in\{1,2| \ell\neq k\}$, to the $k$-th node, is denoted by $\bm{H}_{\ell k}\in\mathbb{C}^{M\times N}$ as well as  $\bm{H}_{kk}\in\mathbb{C}^{M\times N}$ describing the \ac{SI} channel at the $k$-th node.
In light of the above, the received signal at the $k$-th node after processing by the analog \ac{SI} cancellation $\bm{C}_{k}$ can be expressed as
\begin{equation}
\label{eq:received_k}
\bm{y}_{k}  \!=\! \underbrace{\bm{H}_{\ell k}\left(\!\bm{v}_{\ell}s_{\ell} + \bm{w}_{\ell}\!\right)}_\text{Distorted intended signal} \!\!+\! \overbrace{\bm{H}_{kk}\bm{v}_{k}s_{k} - \bm{C}_{k}\bm{v}_{k}s_{k} + \bm{H}_{kk}\bm{w}_{k}}^\text{Cancelled \ac{SI} \& Hardware distortion} + \bm{n}_{k},
\end{equation}
where $\bm{w}_{\ell}\sim\mathcal{CN}\left(\bm{0},\beta\cdot{\rm diag}\left(\bm{v}_{\ell}\bm{v}^{\rm H}_{\ell}\right)\right)$ denotes the nonlinear hardware impairments induced by the $\ell$-th node \cite{OmidTVT18,IimoriCAMSAP17,OmidTWC2018,IimoriICC18}, $\bm{w}_{k}\sim\mathcal{CN}\left(\bm{0},\beta\cdot{\rm diag}\left(\bm{v}_{k}\bm{v}^{\rm H}_{k}\right)\right)$ describes the effect of the self-induced nonlinearity at the $k$-th node, and $\beta$ expresses the hardware distortion level, whereas  $\bm{n}_{k}\sim \mathcal{CN}\left(0,\sigma^2\mathbf{I}_{M}\right)$ denotes the complex \ac{AWGN} vector at the $k$-th receiver.

\subsection[]{Imperfect \ac{CSI} model}

In this subsection, statistical channel models for the communication and \ac{SI} channel matrices ($i.e.,$ $\bm{H}_{\ell k}$ and $\bm{H}_{k k}$) will be described, respectively, while introducing the associated imperfection models relying on the Gauss-Markov theorem \cite{MusavianTVT07,WangTWC07}.

Due to the dominant \ac{LoS} stemming from deterministic close proximity between transmit and receive antennas at the \ac{FD} node, the associated \ac{SI} channel matrix $\bm{H}_{k k}$ can be modeled as the Rician fading channel  \cite{duarte2012experiment}, namely,
\begin{eqnarray}
\bm{H}_{kk} = \sqrt{\frac{\kappa}{1+\kappa}}\bm{H}^{\rm LOS}_{kk}+\sqrt{\frac{1}{1+\kappa}}\bm{H}^{\rm NLOS}_{kk} \:\:\forall k,
\end{eqnarray}
with $\kappa$ denoting the Rician shaping parameter, also referred to as the Rician $K$-factor, which expresses the power contribution of the \ac{LoS} components relative to \ac{NLoS} counterparts, while $\bm{H}^{\rm NLOS}_{kk}$ corresponds to sum of  \ac{NLoS} paths such that each element of $\bm{H}^{\rm NLOS}_{kk}$ follows an \ac{i.i.d.} complex Gaussian variable with zero mean and unit-variance, and the \ac{LoS} component $\bm{H}^{\rm LOS}_{kk}$ can be written as a product of phase array responses $\mathbf{a}_{\rm TX}\left(\theta_{T}\right)$ and $\mathbf{a}_{\rm RX}\left(\theta_{R}\right)$ of the transmit and receive antennas, respectively, that is, 
\begin{eqnarray}
\bm{H}^{\rm LOS}_{kk} = \alpha_k\mathbf{a}^{\rm H}_{\rm RX}\left(\theta_{R}\right)\mathbf{a}_{\rm TX}\left(\theta_{T}\right) \in \mathbb{C}^{M\times N},
\end{eqnarray}
where $\alpha_k$ is a complex gain, $\theta_{T}$ and $\theta_{R}$ denote the \ac{AoD} and \ac{AoA}, respectively, and the associated array responses can be written as 
\begin{eqnarray}
\label{eqn:TXSS}
\mathbf{a}_{\rm TX}\left(\theta_{T}\right)\hspace{-1ex}&=&\hspace{-1ex}\left[\begin{array}{llll}\hspace{-1ex}1 \hspace{-1ex}& e^{j2\pi d {\rm cos}\left(\theta_{T}\right)} \hspace{-1ex}& \hspace{-1ex}\cdots \hspace{-1ex}& e^{j2\pi d\left(N-1\right) {\rm cos}\left(\theta_{T}\right)}\end{array}\right],\\
\label{eqn:RXSS}
\mathbf{a}_{\rm RX}\left(\theta_{R}\right) \hspace{-1ex} &=& \hspace{-1ex}\left[\begin{array}{llll}\hspace{-1ex}1 \hspace{-1ex}& e^{j2\pi d {\rm cos}\left(\theta_{R}\right)}\hspace{-1ex} & \hspace{-1ex}\cdots \hspace{-1ex}& e^{j2\pi d\left(M-1\right) {\rm cos}\left(\theta_{R}\right)}\end{array}\right],
\end{eqnarray}
where we assume that both \ac{FD} nodes are equiped with   \ac{ULA} with half-wavelength antenna spacing $d$.

Given the above, it is assumed hereafter that \ac{CSI} knowledge of the communication and \ac{SI} channel matrices is partially available at the nodes, so that the corresponding imperfect \ac{CSI} can be expressed via the Gauss-Markov uncertainty model as 
\vspace{-1ex}
\begin{eqnarray}
\label{eqn:ChannelModel_Com}
\bm{H}_{k\ell} \!\!\!&=&\!\!\! \sqrt{\left(1-\tau^2_{k\ell}\right)}\hat{\bm{H}}_{k\ell} + \tau_{k\ell}\bm{E}_{k\ell},\\
\label{eqn:ChannelModel_SI}
\bm{H}_{kk} \!\!\!&=&\!\!\! \sqrt{\frac{q_{kk}\kappa}{1+\kappa}}\bm{H}^{\rm LOS}_{kk}\!+\!\sqrt{\frac{q_{kk}}{1\!+\!\kappa}\!}\big(\sqrt{\!1 \!-\!\tau^2_{kk}}\hat{\bm{H}}^{\rm NLOS}_{kk}\!\! + \tau_{kk}\bm{E}_{kk}\big)\nonumber\\
\!\!\!&=&\!\!\! \hat{\bm{H}}_{kk} + \tau^{\prime}_{kk}\bm{E}_{kk},
\end{eqnarray}
where $\tau_{ij},\:i,j\in\{k,\ell\}$ denote parameters of the \ac{CSI} accuracy, $\bm{E}_{ij}$ and $\bm{E}_{ii}$ are the channel estimation error matrices with its elements following \ac{i.i.d.} $\mathcal{CN}\left(0,q_{ij}\right)$ and \ac{i.i.d.} $\mathcal{CN}\left(0,1\right)$, respectively, where $q_{ij}$ and $q_{ii}$ are the path loss gains of the channels $\bm{H}_{ij}$ and $\bm{H}_{ii}$.

Notice that in equation \eqref{eqn:ChannelModel_SI}, we implicitly define the known \ac{SI} components $\hat{\bm{H}}_{ii}$ and the scaled \ac{SI} \ac{CSI} accuracy $\tau^{\prime}_{ii}$ for later convenience, which are, respectively, given by
\begin{eqnarray}
\hat{\bm{H}}_{ii}&\triangleq&\sqrt{\frac{q_{ii}\kappa}{1+\kappa}}\bm{H}^{\rm LOS}_{ii}\!+\!\sqrt{\frac{q_{ii}(1 -\tau^2_{ii})}{1+\kappa}}\hat{\bm{H}}^{\rm NLOS}_{ii},\\
\tau^{\prime}_{ii}&\triangleq&\sqrt{\frac{q_{ii}\cdot\tau^2_{ii}}{1\!+\!\kappa}\!}.
\end{eqnarray}\vspace{-2ex}

Furthermore, we considered in equation \eqref{eqn:ChannelModel_SI} that perfect (or considerably accurate) knowledge of the \ac{LoS} component of the \ac{SI} channel is available due to the deterministic (or much slowly-varying) nature of this channel component \cite{IimoriTWC19}, implying that only part of the \ac{NLoS} components possesses uncertainties.

\subsection{Signal model}

Taking into account the imperfect \ac{CSI} model described in the previous section, plugging equation \eqref{eqn:ChannelModel_Com} and \eqref{eqn:ChannelModel_SI} into the received signal expression given in equation \eqref{eq:received_k} yields
\begin{eqnarray}
\label{eq:received_k_mod}
\bm{y}_{k} \hspace{-4ex}&&=\! \overbrace{\sqrt{1\!-\!\tau^2_{\ell k}}\hat{\bm{H}}_{\ell k}\!\left(\!\bm{v}_{\ell}s_{\ell} \!+\! \bm{w}_{\ell}\!\right)\! +\! \tau_{\ell k}\bm{E}_{\ell k}\left(\!\bm{v}_{\ell}s_{\ell}\!+ \!\bm{w}_{\ell}\!\right)}^\text{Intended signal with \ac{CSI} \& hardware imperfection}\! \\
 \hspace{-4ex}&& \hspace{2ex} + \! \underbrace{\tilde{\bm{H}}_{kk}\bm{v}_{k}s_{k} + \tau^{\prime}_{kk}\bm{E}_{kk}\left(\bm{v}_{k}s_{k}\!+\!\bm{w}_{k}\right)\!+\! \hat{\bm{H}}_{kk}\bm{w}_{k}}_\text{Residual \ac{SI} with \ac{CSI} \& hardware imperfection} + \bm{n}_{k},\nonumber
\end{eqnarray}
with $\tilde{\bm{H}}_{kk}\triangleq\hat{\bm{H}}_{kk} - \bm{C}_{k}$ implicitly being defined.

From equation \eqref{eq:received_k_mod}, the averaged \ac{SINR} and corresponding \ac{MSE} at the $k$-th node can be written in a closed-form expression, respectively, as
\begin{eqnarray}
\label{eq:SINR}
\gamma_{k} &=& \frac{P_{\mathrm{Com},k}}{\Sigma_{k}},\\
\label{eq:MSE}
\varepsilon_{k} &=& \Exp{\!\left(s_{\ell}-\hat{s}_{\ell}\right)\left(s_{\ell}-\hat{s}_{\ell}\right)^{*}} = \frac{1}{\gamma_{k}},
\end{eqnarray}
where the power of the intended signal and interference-plus-noise components can be respectively expressed as
\begin{eqnarray}
P_{\mathrm{Com},k}\!\!\!\!&\triangleq&\!\!\!\! \left(1\!-\!\tau^2_{\ell k}\right)\bm{u}_{k}\hat{\bm{H}}_{\ell k}\bm{v}_{\ell}\bm{v}^{\rm H}_{\ell}\hat{\bm{H}}^{\rm H}_{\ell k}\bm{u}^{\rm H}_{k},\\
\label{eq:SIGMA_IDN}
\Sigma_{k}  \!\!\!\!&\triangleq&\!\!\!\!   \bm{u}_{k}\tilde{\bm{H}}_{kk}\bm{v}_{k}\bm{v}^{\rm H}_{k}\tilde{\bm{H}}^{\rm H}_{kk}\bm{u}^{\rm H}_{k}  + {\tau^{\prime}}^2_{kk}\norm{\bm{v}_k}^2\left(1+\beta\right)\nonumber\\
\!\!\!\!&&\!\!\!\! \!\!\!\!+ \beta\bm{u}_{k}\hat{\bm{H}}_{k k}\diag{\bm{v}_{k}\bm{v}^{\rm H}_{k}}\hat{\bm{H}}^{\rm H}_{k k}\bm{u}^{\rm H}_{k}  \!+\! q_{\ell k}\tau^2_{\ell k}\norm{\bm{v}_\ell}^2\!(1\!+\!\beta) \nonumber\\
\!\!\!\!&&\!\!\!\!\!\!\!\!+ \beta\left(1-\tau^2_{\ell k}\right)\bm{u}_{k}\hat{\bm{H}}_{\ell k}\diag{\bm{v}_{\ell}\bm{v}^{\rm H}_{\ell}}\hat{\bm{H}}^{\rm H}_{\ell k}\bm{u}^{\rm H}_{k}+ \sigma^2,
\end{eqnarray}
where the identity $\Exp{\bm{H}\bm{A}\bm{H}^{\rm H}} = \sigma^2\tr{\bm{A}}\mathbf{I}$ with each element of $\bm{H}$ follows \ac{i.i.d.} $\mathcal{CN}\left(0,\sigma^2\right)$ is leveraged.

Furthermore, for later convenience, the covariance of residual distorted \ac{SI} after the analog cancellation can be written as
\begin{eqnarray}
\hspace{-1.7ex}&&\hspace{-4ex}\mathbf{\Phi}_k\triangleq\mathbb{V}\left[\tilde{\bm{H}}_{kk}\bm{v}_{k}s_{k} + \tau^{\prime}_{kk}\bm{E}_{kk}\left(\bm{v}_{k}s_{k}\!+\!\bm{w}_{k}\right)\!+\! \hat{\bm{H}}_{kk}\bm{w}_{k}\right]\\
\hspace{-1.7ex}&=&\!\!\!\!\!\mathbb{V}\left[\tilde{\bm{H}}_{kk}\bm{v}_{k}s_{k}\right] +  \mathbb{V}\left[\tau^{\prime}_{kk}\bm{E}_{kk}\left(\bm{v}_{k}s_{k}\!+\!\bm{w}_{k}\right)\right] + \mathbb{V}\left[\hat{\bm{H}}_{kk}\bm{w}_{k}\right]\nonumber\\
\hspace{-1.7ex}&=&\!\!\!\! \!\!\tilde{\bm{H}}_{kk}\bm{v}_{k}\bm{v}^{\rm H}_{k}\tilde{\bm{H}}^{\rm H}_{kk}\!\!+\! {\tau^{\prime}}^2_{\!\!\!kk}\!\norm{\bm{v}_k}^{\!2}\!\!(1\!+\!\beta)\mathbf{I} \!+\! \beta\hat{\bm{H}}_{k k}\text{diag}(\bm{v}_{k}\bm{v}^{\rm H}_{k})\hat{\bm{H}}^{\rm H}_{k k}.\nonumber
\end{eqnarray}

Please note from the above that the diagonal elements of $\mathbf{\Phi}_k$ describes the total average \ac{SI} power at each digital thread at the $k$-th receiver, which therefore need to be sufficiently attenuated before processing by the \ac{RF} chain so as to avoid saturation of \ac{LNA} while maintaining the operation point of \ac{LNA} sufficiently high in terms of energy efficiency of \ac{RF} circuits.
To elaborate, the tunable radio components ($i.e.,$ $\bm{v}_k$, $\bm{u}_k$, and $\bm{C}_k$) need to be designed such that 
the $m$-th diagonal element $[\mathbf{\Phi}_k]_{mm}$ satisfies $[\mathbf{\Phi}_k]_{mm}\leq \varepsilon_{k,m}$ with $\varepsilon_{k,m}$ denoting a power level requirement such that the total received signal $\bm{y}_{k}$ enjoys linearity of the dynamic range at the receiver side.

 \section[]{Proposed \ac{SI} cancellation design}
 \label{sec:proposed}
 
Taking into account the fact that maximizing \ac{SINR} at each user  corresponds to minimizing the associated \ac{MSE} as shown in equation \eqref{eq:SINR} and \eqref{eq:MSE}, in this section we shall hereafter consider the following sum \ac{SINR} maximization problem subject to the maximum transmit power constraint at each user as well as the residual \ac{SI} power level constraints at each \ac{RF} thread of the receiver, which can be expressed as
\vspace{-1ex}
\begin{subequations}
\label{eqn:Equ_OP_FP}
\begin{eqnarray}
\!\!\!\!\max_{\bm{v}_{k},\bm{v}_{\ell},\bm{u}_{k},\bm{u}_{\ell}}&&\!\!\!\!\!\!\!\!  g(\bm{v}_{k},\bm{v}_{\ell},\bm{u}_{k},\bm{u}_{\ell})\triangleq\sum^{2}_{k=1}\gamma_{k}
\label{eqn:Equ_OP_FP_Obj}\\
{\rm s.t.}&&\!\!\!\!\!\!\!\!\!\! \norm{\bm{v}_{k}}^2 \leq P_{k}, \forall k\\
\label{eqn:SI_level_analog}
&&\!\!\!\!\!\!\!\!\!\! [\mathbf{\Phi}_k]_{mm}\leq \varepsilon_{k,m}, \forall k, m\in\{1,2,\ldots,M\}.
\end{eqnarray}
\end{subequations}
\vspace{-2ex}

One may readily notice that the optimization problem given in equation \eqref{eqn:Equ_OP_FP} is an intractable non-convex problem due to not only the non-convexity of the \ac{SINR} expressions in equation \eqref{eq:SINR} but also the coupling effect between the variables ($i.e.,$ $\bm{v}_{k},\bm{v}_{\ell},\bm{u}_{k}$ and $\bm{u}_{\ell}$).
Aiming at relaxing this difficulty while taking advantage of the optimality of linear \ac{MMSE} receiving filter in case that both the intended signal and the effective interfering signals can be treated as Gaussian \cite{Negro2010ITA}, we propose a type of the alternating optimization framework in conjunction with the non-monotone algorithmic design \cite{LiNIPS15,QuanmingIJCAI17}.
To this end, the normalized \ac{MMSE} receiving filters at the $k$-th and $\ell$-th node can be, respectively, written in a closed-form expression as
\begin{subequations}
\label{eqn:uk_MMSE}
\begin{equation}
\!\bm{u}_{k} \!=\! \tfrac{\bm{v}^\mathrm{H}_{\ell}\hat{\bm{H}}^\mathrm{H}_{\ell k}\big(\!\bm{H}_{\bm{u}_{k}}\bm{H}^\mathrm{H}_{\bm{u}_{k}}\!\!+\! {\sigma^{\prime}}^2_{\!\!u_{k}}\mathbf{I}_{M} \!+\!(1\!-\!\tau^2_{\ell k})\hat{\bm{H}}_{\ell k}\bm{v}_{\ell}\bm{v}^{\rm H}_{\ell}\hat{\bm{H}}^{\rm H}_{\ell k}\big)^{\!-1}}{\big\|\bm{v}^\mathrm{H}_{\ell}\hat{\bm{H}}^\mathrm{H}_{\ell k}\big(\!\bm{H}_{\bm{u}_{k}}\bm{H}^\mathrm{H}_{\bm{u}_{k}}\!\!+\! {\sigma^{\prime}}^2_{\!\!u_{k}}\mathbf{I}_{M} \!+\!(1\!-\!\tau^2_{\ell k})\hat{\bm{H}}_{\ell k}\bm{v}_{\ell}\bm{v}^{\rm H}_{\ell}\hat{\bm{H}}^{\rm H}_{\ell k}\big)^{-1}\big\|_2},
\end{equation}
\begin{equation}
\bm{u}_{\ell} \!=\! \tfrac{\bm{v}^\mathrm{H}_{k}\hat{\bm{H}}^\mathrm{H}_{k\ell}\big(\!\bm{H}_{\bm{u}_{\ell}}\bm{H}^\mathrm{H}_{\bm{u}_{\ell}}\!\!+\! {\sigma^{\prime}}^2_{\!\!u_{\ell}}\mathbf{I}_{M} \!+\!(1\!-\!\tau^2_{k\ell})\!\hat{\bm{H}}_{k\ell}\bm{v}_{k}\bm{v}^{\rm H}_{k}\hat{\bm{H}}^{\rm H}_{k\ell}\big)^{-1}}{\big\|\bm{v}^\mathrm{H}_{k}\hat{\bm{H}}^\mathrm{H}_{k\ell}\big(\!\bm{H}_{\bm{u}_{\ell}}\bm{H}^\mathrm{H}_{\bm{u}_{\ell}}\!\!+\! {\sigma^{\prime}}^2_{\!\!u_{\ell}}\mathbf{I}_{M} \!+\!(1\!-\!\tau^2_{k\ell})\!\hat{\bm{H}}_{k\ell}\bm{v}_{k}\bm{v}^{\rm H}_{k}\hat{\bm{H}}^{\rm H}_{k\ell}\big)^{-1}\big\|_2},
\end{equation}
\end{subequations}
where the effective interfering channels $\bm{H}_{\bm{u}_{k}}$ and $\bm{H}_{\bm{u}_{\ell}}$ are given in equation \eqref{eq:Hu}
\begin{figure*}[t]
\begin{subequations}
\label{eq:Hu}
\begin{equation}
\label{eqn:Huk}
\bm{H}_{\bm{u}_{k}} \triangleq
\left[
\tilde{\bm{H}}_{\!kk}\bm{v}_{k},
\beta\hat{\bm{H}}_{\!kk}\bm{\Gamma}^{N}_{1} \bm{v}_{k},
 \beta(1\!-\!\tau^2_{\ell k})\hat{\bm{H}}_{\!\ell k}\bm{\Gamma}^{N}_{1}\!\bm{v}_{\ell},
 \cdots,
 \beta\hat{\bm{H}}_{\!kk}\bm{\Gamma}^{N}_{N}\bm{v}_{k},
 \beta(1\!-\!\tau^2_{\ell k})\hat{\bm{H}}_{\!\ell k}\bm{\Gamma}^{N}_{N}\bm{v}_{\ell}
\right]
\!\!
\end{equation}
\begin{equation}
\label{eqn:Hul}
\bm{H}_{\bm{u}_{\ell}} \triangleq
\left[
\tilde{\bm{H}}_{\!\ell\ell}\bm{v}_{\ell},
 \beta\hat{\bm{H}}_{\!\ell\ell}\bm{\Gamma}^{N}_{1}\bm{v}_{\ell},
 \beta(1\!-\!\tau^2_{k\ell})\hat{\bm{H}}_{k\ell}\bm{\Gamma}^{N}_{1}\bm{v}_{k},
 \cdots,
\beta\hat{\bm{H}}_{\ell\ell}\bm{\Gamma}^{N}_{N}\bm{v}_{\ell},
 \beta(1\!-\!\tau^2_{k\ell})\hat{\bm{H}}_{k\ell}\bm{\Gamma}^{N}_{N}\bm{v}_{k}
\right]\!\!
\vspace{-1ex}
\end{equation}
\end{subequations}
\setcounter{equation}{21}
\begin{equation}
\label{eqn:gradient}
\nabla f\left(\bm{v}_{k}\right) = \left(\frac{\partial P_{\mathrm{Com},k}}{\partial \bm{v}^{*}_{k}}\Sigma_{k} - \frac{\partial \Sigma_{k}}{\partial \bm{v}^{*}_{k}}  P_{\mathrm{Com},k} \right)\frac{1}{\Sigma^2_{k}} + \left(\frac{\partial P_{\mathrm{Com},\ell}}{\partial \bm{v}^{*}_{k}}\Sigma_{\ell} - \frac{\partial \Sigma_{\ell}}{\partial \bm{v}^{*}_{k}}  P_{\mathrm{Com},\ell} \right)\frac{1}{\Sigma^2_{\ell}},
\end{equation}
\setcounter{equation}{18}
\hrule
\vspace{-1ex}
\end{figure*}
with $\bm{\Gamma}^{N}_{i}\in\mathbb{R}^{N\times N}$ being an all-zero matrix except for its $i$-th diagonal position equal to $1$. 

Given fixed receiving filters calculated by equation \eqref{eqn:uk_MMSE}, the optimization problem \eqref{eqn:Equ_OP_FP} can then be reduced to 
\begin{subequations}
\label{eqn:Equ_OP_FP_mod1}
\begin{eqnarray}
\label{eqn:Equ_OP_FP_obj_mod1}
\!\!\!\!\max_{\bm{v}_{k},\bm{v}_{\ell}}&&\!\!\!\!\!\!\!\! f\left(\bm{v}_{k},\bm{v}_{\ell}\right)\triangleq \sum^{2}_{k=1}\gamma_{k}\\
\label{eqn:PowerConst}
{\rm s.t.}&&\!\!\!\!\!\!\!\!\!\! \norm{\bm{v}_{k}}^2 \leq P_{k}, \forall k\\
\label{eqn:SI_level_analog_1}
&&\!\!\!\!\!\!\!\!\!\! [\mathbf{\Phi}_k]_{mm}\leq \varepsilon_{k,m}, \forall k, m\in\{1,2,\ldots,M\}.
\end{eqnarray}
\end{subequations}

Notice that the residual \ac{SI} power constraints \eqref{eqn:SI_level_analog_1} are independent from the receiving filters $\bm{u}_{k}$ and $\bm{u}_{\ell}$ since it is needed to be satisfied before processing by the \ac{LNA} and \ac{ADC}, indicating that the precoding vectors must be designed so as to satisfy both the transmit power constraint and \eqref{eqn:SI_level_analog_1} simultaneously.
In order to tackle the non-convexity of the objective function \eqref{eqn:Equ_OP_FP_obj_mod1} while enjoying the convex sets \eqref{eqn:PowerConst} and \eqref{eqn:SI_level_analog_1}, the \ac{FD} communication literature \cite{OmidTWC2018, BDaySAC2012,IimoriTWC19} as well as recent machine learning research works \cite{LiNIPS15,QuanmingIJCAI17} jointly motivate us to propose an alternating non-monotone gradient projection method described as follows. 

A fundamental algorithmic framework of gradient projection methods can be written as 
\vspace{-1ex}
\begin{eqnarray}
\label{eqn:projectedvk}
\breve{\bm{v}}^{[t]}_{k} \!\!&=&\!\! \mathcal{P}\Big\{\overbrace{\bm{v}^{[t]}_{k} + \delta^{[t]} \nabla f\left(\bm{v}_{k}\right)}^{\triangleq \bar{\bm{v}}_{k}}\Big\}\\
\bm{v}^{[t+1]}_{k} \!\!&=&\!\! \bm{v}^{[t]}_{k} + \rho^{[t]}\left(\breve{\bm{v}}^{[t]}_{k}- \bm{v}^{[t]}_{k}\right)
\label{eqn:newvk}
\end{eqnarray}
where $\mathcal{P}\left\{\cdot\right\}$ denotes the projection operator that computes the closest point (in Euclidean sense) from a current estimate towards the feasible set, which is described later in details, $\delta$ and $\rho$ are the intensity parameters for the projection and correction steps, and $\nabla f\left(\bm{v}_{k}\right)$ is the gradient of $f\left(\bm{v}_{k}\right)$, which can be written in a closed-form expression as shown above in equation \eqref{eqn:gradient}, with  
\setcounter{equation}{22}
\begin{subequations}
\begin{eqnarray}
\frac{\partial P_{\mathrm{Com},k}}{\partial \bm{v}^{*}_{k}} \!\!\!&=&\!\!\! \bm{0}_{N\times 1},\\
\frac{\partial P_{\mathrm{Com},\ell}}{\partial \bm{v}^{*}_{k}} \!\!\!&=&\!\!\! \left(1\!-\!\tau^2_{k\ell}\right)\hat{\bm{H}}^{\rm H}_{k\ell}\bm{u}^{\rm H}_{\ell}\bm{u}_{\ell}\hat{\bm{H}}_{k\ell}\bm{v}_{k},\\
\frac{\partial \Sigma_{k}}{\partial \bm{v}^{*}_{k}} \!\!\!&=&\!\!\!  \tilde{\bm{H}}^{\rm H}_{kk}\bm{u}^{\rm H}_{k}\bm{u}_{k}\tilde{\bm{H}}_{kk}\bm{v}_{k}+ {\tau^{\prime}}^2_{kk}\left(1+\beta\right)\bm{v}_{k}\nonumber\\[-1.5ex]
&&\hspace{2ex} + \beta\sum^{N}_{i=1}\bm{\Gamma}^{N}_{i}\hat{\bm{H}}^{\rm H}_{k k}\bm{u}^{\rm H}_{k}\bm{u}_{k}\hat{\bm{H}}_{k k}\bm{\Gamma}^{N}_{i}\bm{v}_{k},\\[-1ex]
\frac{\partial \Sigma_{\ell}}{\partial \bm{v}^{*}_{k}} \!\!\!&=&\!\!\!  \beta(1-\tau^2_{k\ell})\sum^{N}_{i=1}\bm{\Gamma}^{N}_{i}\hat{\bm{H}}^{\rm H}_{k \ell}\bm{u}^{\rm H}_{\ell}\bm{u}_{\ell}\hat{\bm{H}}_{k\ell}\bm{\Gamma}^{N}_{i}\bm{v}_{k}\nonumber\\[-1ex]
&&\hspace{15ex} + q_{k\ell}\tau^2_{k\ell}(1+\beta)\bm{v}_{k}.
\end{eqnarray}
\end{subequations}

\begin{algorithm}[t!]
\hrulefill
\begin{algorithmic}[1]
\vspace{-0.5ex}
\Statex {\bf{Input:}} $P_\mathrm{max},\hat{\bm{H}}_{kk},\hat{\bm{H}}_{\ell k}$, $\bm{C}_{k},\forall k$, $c$
\Statex {\bf{Output:}} $\bm{v}_{k},\bm{v}_{\ell},\bm{u}_{k},\bm{u}_{\ell}$
\vspace{-1.5ex}
\Statex \hspace{-4ex}\hrulefill
\State   Set iteration number $t=0$ and $P_{k}=P_\mathrm{max}\: \forall k$.
\vspace{0.2ex}
\State  Obtain initial $\bm{v}^{[t]}_{k}\: \forall k$ via Gauss. random init. \cite{IimoriWPNC18,ShenTWC2010}.
\vspace{0.2ex}
\Repeat\vspace{0.2ex}
\State  $t=t+1$.
\vspace{0.2ex}
\State  Compute $\bm{u}^{[t]}_{k}\:\forall k$ from \eqref{eqn:uk_MMSE}.
\vspace{0.2ex}
\State  Get $\bar{\bm{v}}^{[t-1]}_{k}$ according to \eqref{eqn:projectedvk} and \eqref{eqn:gradient}.
\vspace{0.2ex}
\State Project $\bar{\bm{v}}^{[t-1]}_{k}$ onto \eqref{eqn:PowerConst} and \eqref{eqn:SI_level_analog_1}.
\vspace{0.2ex}
\State Obtain $\bm{v}^{[t]}_{k}$ from  \eqref{eqn:newvk}.
\vspace{0.2ex}
\State {\small$\delta \!=\! |g(\bm{v}^{[t]}_{k},\bm{v}^{[t]}_{\ell},\bm{u}^{[t]}_{k},\bm{u}^{[t]}_{\ell}) - g(\bm{v}^{[t-1]}_{k},\bm{v}^{[t-1]}_{\ell},\bm{u}^{[t-1]}_{k},\bm{u}^{[t-1]}_{\ell})|$}.
\vspace{0.5ex}
\State $\Delta^{[t]} = \argmax_{t^\prime = \{\text{max}(1,t-c),\ldots,t\}}g\big(\bm{v}^{[t^\prime]}_{k},\bm{v}^{[t^\prime]}_{\ell},\bm{u}^{[t^\prime]}_{k},\bm{u}^{[t^\prime]}_{\ell}\big)$.
\vspace{0.2ex}
\If{$\delta<10^{-6}$ or ($\Delta^{[t]}=t-c$)}
\State $\bm{u}^{[t]}_{k} = \bm{u}^{[\Delta^{[t]}]}_{k}\:\forall k$.
\vspace{0.2ex}
\State $\bm{v}^{[t]}_{k} = \bm{v}^{[\Delta^{[t]}]}_{k}\:\forall k$.
\vspace{0.2ex}
\State break.
\EndIf
\Until {\rm reach maximum iterations}
\caption[]{:\\\small ALTernating Non-Monotone GrAdient Projection (ALTnmGAP)}
\label{alg:main}
\end{algorithmic}
\end{algorithm}
\setlength{\textfloatsep}{0pt}

Following \cite{OmidTWC2018}, we adopt the Armijo rule for updating the step size parameters $\delta$ and $\rho$, which can be expressed as
\begin{equation}
\label{eqn:ArmijoStep}
f\!\big(\bm{v}^{[n+1]}_{k}\big)\!-\!f\!\big(\bm{v}^{[n]}_{k}\big)\!\geq\! \iota \nu^{m} \tr{\!\nabla f^\mathrm{H}\big(\bm{v}_{k}\big)\big(\breve{\bm{v}}^{[n]}_{k}\!-\! \bm{v}^{[n]}_{k}\big)\!},
\end{equation}
with $m$ denoting  the minimal positive integer satisfying the above inequality with $\nu=0.5$, $\iota=0.1$, $\delta=1$ and $\rho^{[n]}=\nu^m$.

In light of the above, the gradient step can be calculated according to equations \eqref{eqn:newvk} and \eqref{eqn:ArmijoStep}.
In what follows, we describe how to perform the projection operation onto the feasible set given in equation \eqref{eqn:PowerConst}, \eqref{eqn:SI_level_analog_1}, and \eqref{eqn:projectedvk}.
Since the feasible region characterized by equation \eqref{eqn:PowerConst} and \eqref{eqn:SI_level_analog_1} is a convex set, we readily obtain
\begin{subequations}
\begin{eqnarray}
\!\!\!\!\!\!\!\!\!\!\!\mathcal{P}\left\{\bar{\bm{v}}_{k}\right\} \!=\!\!\!\!\! &\argmin_{\bm{z}}& \norm{\bar{\bm{v}}_{k} - \bm{z}}\\[-1ex]
&{\rm s.t.}&\!\!\!\! \norm{\bm{z}}^2 \leq P_{k}\\
&&\!\!\!\! [\mathbf{\Phi}_k]_{mm}\leq \varepsilon_{k,m}, m\in\{1,2,\ldots,M\},\\[-4ex]\nonumber
\end{eqnarray}
\end{subequations}
where one may notice that the above problem is an optimization type of convex \acp{QCQP} which can be efficiently solved not only by the well-known interior point methods\footnote{Note that interior point methods are widely available, $i.e.,$ CVX \cite{CVX2017} in Matlab, CVXPY \cite{diamond2016cvxpy} in Python, and Convex.jl \cite{Udell2014CVXjl} in Julia.}  \cite{AndersenMOSEK00} but also by leveraging recently proposed specialized solvers such as \cite{AdachiMP2019}.

To conclude this section, we offer a summary of the proposed alternating non-monotone gradient projection method here developed in the form of a pseudo code in Algorithm \ref{alg:main}, where the non-monotonicity of the gradient step is also algorithmically explained.
The authors kindly refer an interested reader to \cite{LiNIPS15,QuanmingIJCAI17} for convergence guarantee of the non-monotone gradient methods, which can be extended to the proposed method and is omitted due to the space limit on the page length.

\section{Simulation Results}
\label{sec:results}

In this section, we evaluate via software simulations the performance of the proposed method in comparison with state-of-the-art methods in terms of the achievable throughput as well as the residual \ac{SI} at the receiver analog domain.
Following the related works \cite{Alexandropoulos2017,IimoriTWC19}, we compare our proposed method with recent state-of-the-art algorithms ($i.e.,$ AltDRQ \cite{IimoriTWC19} and AltRQSpl \cite{Alexandropoulos2017}).
The simulation setups are as follows.

In order to be in line with prior works such as \cite{IimoriTWC19,Alexandropoulos2017, duarte2012experiment}, the numbers of transmit and receive antennas at each node are assumed to be $M=N=4$ and maximum transmit power $P_{\rm max}$ is limited to $P_{\rm max}=20$ dBm, where the noise floor is set to $-90$ dBm.
Also, the number of analog cancellation taps $N_{\rm tap}$ is equivalent to $8$, where one may notice that this is $50$\% reduction in the number of elements in the analog cancellation matrix $\bm{C}_{k}$ in comparison with \cite{bharadia2013fullduplex, Everett2011Asilomar}.
Aiming at modeling practical imperfect analog \ac{SI} cancellation, it is assumed that each analog tap of $\bm{C}_{k}$ suffers from amplitude imperfection uniformly distributed between $-0.01$dB and $0.01$dB, while the associated phase noise is uniformly distributed between $-0.065^{\circ}$ and $0.065^{\circ}$ \cite{Kolodziej2016TWC}, while the target residual \ac{SI} level is considered to be $\varepsilon_{k,m}=\varepsilon=-47$ dBm for all $k$ and $m$.

Furthermore, the communication channel matrices $\bm{H}_{k\ell}$ and $\bm{H}_{\ell k}$ are assumed to follow block Rayleigh fading channel with $110$dB pathloss, while
the \ac{SI} channels $\bm{H}_{kk}$ and $\bm{H}_{\ell\ell}$ are considered to be modeled as block Rician fading channel with $40$dB pathloss and a $35$dB $K$-factor, where the channel estimation accuracy levels $\tau_{k}$ and $\tau_{\ell}$ are statistically equivalent ($i.e.,$ $\tau_{k}=\tau_{\ell}=\tau$) with its effective range of $\tau\in\{-40,-15\}$ dB.
For the \ac{LoS} components of the \ac{SI} channels, it is assumed that \ac{AoD} $\theta_{\rm T}$ and \ac{AoA} $\theta_{\rm R}$ are uniformly distributed over the phase domain ($i.e.,$ $\{0,2\pi\}$).

In addition to the above, the hardware nonlinearity factors $\beta_{k}\: \forall k$ are assumed to be identical ($i.e.$, $\beta_{k}=\beta_{\ell}=\beta$), and a moderate hardware impairment level is considered $\beta\in-50$ dB \cite{OmidTWC2018}.
The algorithmic parameters such as the memory length $c$ and the maximum number of iterations $t_{\rm max}$ are chosen to be $c=8$ and $t_{\rm max}=50$.

\begin{figure}[t!]
\centering
\includegraphics[width=0.97\columnwidth]{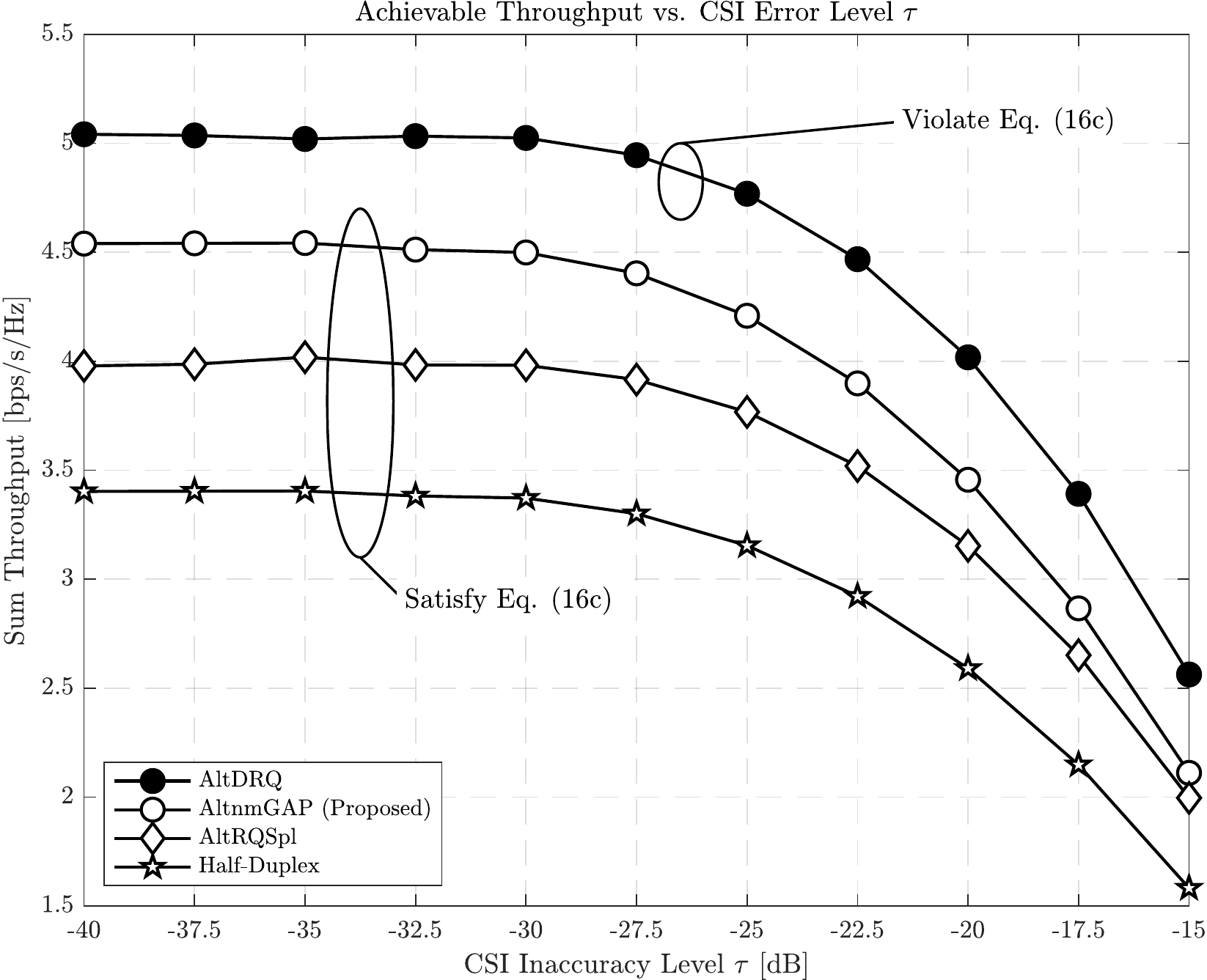}\vspace{-1ex}
\caption[]{Achievable sum-throughput comparisons of the proposed method against the state-of-the-art \ac{SI} cancellation methods as well as the conventional half-duplex as a function of the channel estimation accuracy level $\tau$ with a moderate hardware distortion level $\beta=-50$ dB.}
\label{fig:throughput}
\vspace{2ex}
\centering
\includegraphics[width=0.97\columnwidth]{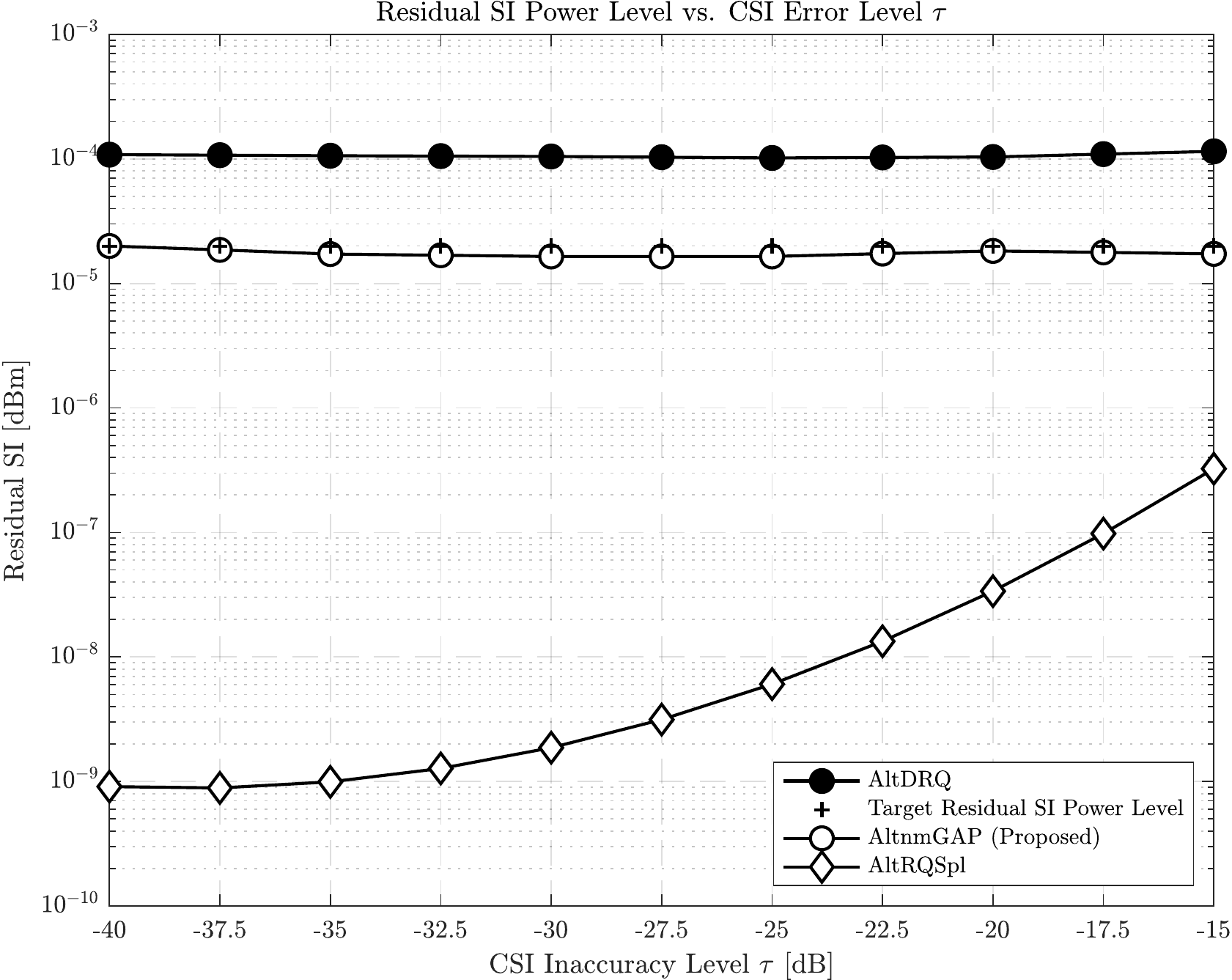}\vspace{-1.5ex}
\caption[]{Residual \ac{SI} power levels after processing by the analog cancellation at the \ac{FD} receiver for different methods ($i.e.,$ the proposed and state-of-the-arts) as a function of the channel estimation accuracy level $\tau$ with a moderate hardware distortion level $\beta=-50$ dB, where the target residual \ac{SI} level is highlighted by the marker ``plus''.}\vspace{1ex}
\label{fig:residualSI}
\end{figure}

Figure \ref{fig:throughput} illustrates the achievable sum-throughput performance of the proposed ALTnmGAP method in comparison with the state-of-the-art methods \cite{Alexandropoulos2017,IimoriTWC19} as a function of the \ac{CSI} accuracy parameter $\tau$ subject to moderate hardware impairment level $\beta=-50$ dB, where the performance corresponding to the half-duplex mode is also offered for the sake of comparison.
As highlighted in the figure, it can be observed that the proposed method can achieve the maximum throughput among the methods satisfying the residual \ac{SI} power level given in equation \eqref{eqn:SI_level_analog}, which can be confirmed in Figure \ref{fig:residualSI}.
The residual \ac{SI} power level after processing by the analog cancellation is shown in Figure \ref{fig:residualSI} as a function of $\tau$, which clearly demonstrates the fact that the proposed method can suppress the residual \ac{SI} at the receiver analog domain below the target over a wide range of \ac{CSI} inaccuracy.
Please note in the figure that the target \ac{SI} residual level is denoted by the marker ``plus'' for the sake of readability.
All in all, one may conclude that the proposed method can be seen as a compromise solution balancing the achievable throughput performance and the residual \ac{SI} level requirement at the receiver analog domain so that the \ac{FD} communications can enjoy the \ac{RF} dynamic range at the receiver.

\section{Conclusion}
\label{sec:conclusion}
\vspace{-0.5ex}
This article studied a bidirectional in-band \ac{FD} \ac{MIMO} system subject to imperfect \ac{CSI}, hardware distortion, and limited analog cancellation capability as well as the \ac{SI} power requirement at the receiver analog domain such that the residual \ac{SI} at the receiver analog domain may not pose saturation of the \ac{LNA}.
An optimization problem aiming at maximizing the sum \ac{SINR} while satisfying the transmit power constraint and the residual \ac{SI} power level requirement is formulated, proposing a new gradient projection based \ac{SI} cancellation mechanism in conjunction with the concept of non-monotonicity. 
Simulation results demonstrated that the proposed method is a compromise solution to the latter problem, which balances the throughput performance and residual \ac{SI} requirements.

\bibliographystyle{IEEEtran}
% Ignore errors thrown by references section before editing
%\bibliography{IEEEabrv,\myreferences,MyListOfPapers}
\bibliography{listofpublications}

\end{document}